\documentclass[aps,pra,preprint,showpacs]{revtex4-1}

\usepackage{graphicx}
\usepackage{dcolumn}
\usepackage{amsmath}
\usepackage{mathrsfs}
\usepackage{amsbsy}
\usepackage{amssymb}
\usepackage{mathtools}
\usepackage{upgreek}

\usepackage[twoside]{rotating}
\usepackage{psfrag}
\usepackage[tight,footnotesize]{subfigure}

\usepackage{stackengine}

\usepackage{color,soul}
\definecolor{LightGray}{gray}{0.8}
\definecolor{Orange}{rgb}{1.0, 0.31, 0.0}
\definecolor{Green}{rgb}{0.3, 1.0, 0.3}
\definecolor{Blue}{rgb}{0.75,0.75,1}

\newcommand{\fig}[1]{Fig.~\ref{#1}}
\newcommand{\figs}[1]{Figs~\ref{#1}}

\newcommand{\bea}{\begin{eqnarray}}
\newcommand{\beal}[1]{\begin{eqnarray}\label{#1}}
\newcommand{\eea}{\end{eqnarray}}
\def\balg#1#2\ealg{\begin{align}\label{#1}#2\end{align}}
\def\balgnl#1\ealgnl{\begin{align*}#1\end{align*}}

\newcommand{\x}{{\mathbf r}}

\newcommand{\E}{{\mathbf E}}
\renewcommand{\H}{{\mathbf H}}

\newcommand{\ur}{{\mathbf e}_r}

\let\originaleqref=\eqref
\renewcommand{\eqref}{Eq.~\originaleqref}

\newcommand{\eqrefs}[1]{Eqs~(\ref{#1})}
\newcommand{\eqrefnt}[1]{(\ref{#1})}

\begin{document}


\title{Monitoring strong coupling in nonlocal plasmonics with
electron spectroscopies}



\author{Grigorios P. Zouros}
\email{zouros@ieee.org}
\affiliation{School of Electrical and
Computer Engineering, National Technical University of Athens, Athens 15773, Greece}
\author{Georgios D. Kolezas}
\affiliation{School of Electrical and
Computer Engineering, National Technical University of Athens, Athens 15773, Greece}
\author{N. Asger Mortensen}
\affiliation{Center for Nano Optics,
University of Southern Denmark, Campusvej 55, DK-5230 Odense M, Denmark}
\affiliation{Danish Institute for Advanced
Study, University of Southern Denmark, Campusvej 55, DK-5230 Odense M, Denmark}
\author{Christos Tserkezis}
\email{ct@mci.sdu.dk}
\affiliation{Center for Nano Optics,
University of Southern Denmark, Campusvej 55, DK-5230 Odense M, Denmark}


\date{\today}

\begin{abstract}
Plasmon--exciton polaritons provide exciting possibilities to control
light--matter interactions at the nanoscale by enabling closer
investigation of quantum optical effects and facilitating novel
technologies based, for instance, on Bose--Einstein condensation and
polaritonic lasing. Nevertheless, observing and visualising polaritons
is challenging, and traditional optical microscopy techniques often
lead to ambiguities regarding the emergence and strength of the
plasmon--exciton coupling. Electron microscopy offers a more robust
means to study and verify the nature of plexcitons, but is still
hindered by instrument limitations and resolution. A simple theoretical
description of electron beam-excited plexcitons is therefore vital
to complement ongoing experimental efforts. Here we apply analytic
solutions for the electron-loss and photon-emission probabilities
to evaluate plasmon--exciton coupling studied either with the recently
adopted technique of electron energy-loss spectroscopy, or with the
so-far unexplored in this context cathodoluminescence spectroscopy.
Foreseeing the necessity to account for quantum corrections in the
plasmonic response, we extend these solutions within the framework
of general nonlocal hydrodynamic descriptions. As a specific example
we study core--shell spherical emitter--molecule hybrids, going beyond
the standard local-response approximation through the hydrodynamic
Drude model for screening and the generalised nonlocal optical
response theory for nonlocal damping. We show that electron
microscopies are extremely powerful in describing the interaction
of emitters with the otherwise weakly excited by optical means
higher-order plasmonic multipoles, a response that survives when
quantum-informed models are considered. Our work provides therefore
both a robust theoretical background and supporting argumentation to
the open quest for improving and further utilising electron microscopies
in strong-coupling nanophotonics.
\end{abstract}


\maketitle

\section{Introduction}
Electron-beam spectroscopies have been rapidly gaining their
well-deserved share of attention in nanophotonics, as they have opened
new pathways for the optical characterisation of state-of-the-art
nanoscale architectures \cite{polman_natmat}. Electron energy-loss
spectroscopy (EELS) has proven time and again efficient in mapping
the localised surface plasmon (LSP) modes of metallic nanoparticles
(NPs), thus offering unique insight into nanoscopic optical
processes \cite{nelayah_natphys3,bosman_nanotech18}, including the	
possibilities to optically excite dark modes in NPs \cite{koh_nn3},
or map plasmons in novel materials such as graphene \cite{eberlein_prb77,
zhou_natnano7}. Of particular importance in this context is the
realisation that EELS can be a more accurate probe for nanoscale
effects of quantum origin \cite{scholl_nature483,raza_nanophotot2},
thus accelerating the growth of quantum plasmonics \cite{tame_natphys9,
zhu_natcom7,bozhevolnyi_nanophot6,fernandez-dominguez_acsphotonics2018}. Complementary to the near-field-oriented
EELS is cathodoluminescence (CL) spectroscopy, which is more efficient
at probing radiative modes excited by subnanometre electron
beams \cite{vesseur_nl7,gomez_njp10,losquin_nl15}. Combining these
two techniques, a richness of information on the response of
nanophotonic architectures can be acquired \cite{kuttge_prb79,
chaturvedi_nn3,yamamoto_nl11,raza_natcom6}. 

Recently, EELS was theoretically proposed \cite{konecna_nn12}, \mbox{\cite{crai_acsphotonics2019}}
and experimentally explored \mbox{\cite{yankovich_arxiv2019}}, \mbox{\cite{bitton_et_al}} as an
alternative technique for monitoring strong coupling in nanophotonics
and visualising the formation of hybrid exciton-polaritons. In particular, EELS was experimentally used to trace the anticrossing of two hybrid modes in truncated nanopyramids coupled to excitons in transition-metal dichalcogenides  \mbox{\cite{yankovich_arxiv2019}},  and in quantum dots coupled to dark bowtie-antenna modes \mbox{\cite{bitton_et_al}},
illustrating how electron spectroscopies have nothing to envy from their optical counterparts, but can in fact be more efficient when dark modes are involved. Inspired by quantum optics, strong coupling is among the most rapidly growing
areas in photonics \cite{torma_rpp78,baranov_acsphot5}, because it
combines the possibility to assess quantum-optical concepts without
the need for extreme laboratory conditions \cite{yoshie_nat432,
dovzhenko_nscale10,ojambati_natcom10,tserkezis_arxiv2019} with the
promise of technological advances in a diversity of areas such as
optical nonlinearities \cite{sanvitto_natmat15}, logic gates and
circuits \cite{liew_prl101}, polariton lasing \cite{kena-cohen_natphot4},
Bose--Einstein condensation \cite{hakala_natphys14}, or via
modification of the properties of matter through polaritonic
chemistry \cite{feist_acsphot5} and enablement of forbidden
transitions \cite{cuartero_acsphot5}. For this reason, a plethora
of designs has been proposed, ranging from planar metallic
films \cite{pockrand_jcp77,bellessa_prl93} and metallic
NP arrangements \cite{zengin_prl114,chikkaraddy_nat535,todisco_nn10,
chatzidakis_jmodopt66} combined with organic molecules, to quantum
dots \cite{santhosh_natcom7} or two-dimensional materials \cite{liu_nl16,
geisler_acsphot6} in nanophotonic cavities. Most of these designs exploit
the tremendous field confinement provided by plasmonics, although dielectric
nanocavities with lower losses are now emerging as attractive
alternatives \cite{wang_nl16,lepeshov_acsami10,tserkezis_prb98,todisco_arxiv2019}.
Nevertheless, despite their different approaches in terms of design and
application, what the vast majority of these works have in common is
the use of optical microscopy as the key analysis technique.

Here we turn to more recent efforts to introduce electron microscopy as a
tool for exploring strong coupling \mbox{\cite{konecna_nn12,crai_acsphotonics2019,yankovich_arxiv2019,bitton_et_al}},
and take them one step further by showing that both EELS and CL can provide
information about the occurrence of hybridisation. Furthermore, anticipating
the fabrication of architectures with even finer geometrical details, we develop the appropriate framework to include quantum effects in the
plasmonic response on the basis of standard or more generalised
hydrodynamic models \cite{kosionis_jpcc116,christensen_nn8,raza_jpcc27,
tserkezis_acsphot5a}, appropriately extending very recent theoretical descriptions of classical strong coupling \mbox{\cite{crai_acsphotonics2019}}. While the composites studied here are relatively unrealistic with modern technology, and somehow simplified in terms of design, they can be described by exact analytic solutions which can be used for benchmarking any computational schemes designed to describe more realistic examples. Focusing on core--shell NPs, we provide analytic
solutions for the electron energy-loss (EEL) and photon-emission (PE)
probabilities, which are valuable, not only for obtaining a clear physical
interpretation, but also for benchmarking more elaborate designs. As an
illustrative example, we show that the spectral anticrossing anticipated
for Ag nanospheres covered by, or encapsulating an excitonic layer, can be
efficiently traced in EEL and PE probabilities. Such tools can be advantageous
when the excitons couple to the dominant dipolar plasmon mode, but even
more so in the case of higher-order multipoles, whose linewidths might be
better comparable to those of the excitons, but whose prevailing non-radiative
nature makes their exploration with optical microscopies problematic. Recent experiments \mbox{\cite{raza_natcom6}} have shown that higher-order multipole modes indeed contribute to the EELS signal of ultrasmall nanospheres, with sensitivity that goes way beyond the capabilities of optical spectroscopies.
We thus believe that our work will provide additional motivation to further
invest in exploring strong coupling with electron microscopy.

\section{Analytic solutions}
Let us first describe the general framework for investigating
plasmon--exciton coupling in the local-response approximation (LRA) 
in the case of spherical NPs. Typically, the LRA regime corresponds
to NP radii larger than $\sim 20$\;nm, for which nonlocal effects are
not relevant \cite{tserkezis_prb98}. The plexcitonic configuration
employed here is based on the core--shell geometry, with either a
plasmonic sphere of radius $R_{1}$ covered by an excitonic shell of
outer radius $R_{2}$, as shown in \fig{Fig1}(a), or a plasmonic shell
(outer radius $R_{2}$) encapsulating an excitonic core (radius $R_{1}$),
as shown in \fig{Fig1}(c). In either case, the excitation is a 
swift electron of velocity $\nu$ travelling at a distance $d$ --- the
impact parameter --- from the NP centre (taken as the coordinate
origin). In our calculations we set $d=70$~nm and $\nu=0.69 \; c$,
$c$ being the velocity of light in vacuum. Neglecting relativistic effects, the latter corresponds to a kinetic energy of $\sim$120\,keV.
The relative permittivity
$\varepsilon_{\mathrm{m}}$ of the plasmonic component as a function of
angular frequency $\omega$ follows a Drude model \cite{Bohren_Wiley1983}, i.e., $\varepsilon_{\mathrm{m}}(\omega) = \varepsilon_{\infty}(\omega) -
\omega_\mathrm{p}^{2} / [\omega(\omega - \mathrm{i} \gamma_{\mathrm{m}})]$
(throughout this paper we assume an $\exp (\mathrm{i} \omega t)$ time
dependence of the fields), where $\varepsilon_{\infty}$ accounts
for interband transitions, $\omega_{\mathrm{p}}$ is the plasma
frequency, and $\gamma_{\mathrm{m}}$ is the damping rate in the
metal. In this study we employ Ag, described by $\varepsilon_{\infty}
= 5$ \cite{yang_prb91}, $\hbar \omega_{\mathrm{p}} = 8.99$\;eV, and
$\hbar \gamma_{\mathrm{m}} = 0.025$\;eV \cite{raza_jpcc27}.
These values provide a good Drude fit  of the experimental data by Johnson and Christy \mbox{\cite{joh_chr_72}} in the free-electron regime, while for 
$\varepsilon_{\infty}$ we use the value given in Ref.~\mbox{\cite{yang_prb91}}, to keep
it constant for simplicity. The
relative permittivity $\varepsilon_{\mathrm{e}}$ of the excitonic
material is modeled by a Drude--Lorentz model as $\varepsilon_{\mathrm{e}}
(\omega) = 1 - f \omega_{\mathrm{e}}^{2} / [\omega(\omega - \mathrm{i}
\gamma_{\mathrm{x}}) -\omega_{\mathrm{e}}^{2}]$, with $\hbar
\omega_{\mathrm{e}} = 2.7$\;eV, $\hbar \gamma_{\mathrm{x}} = 0.052$\;eV,
and reduced oscillator strength $f=0.02$ \cite{fofang_nl8,tserkezis_acsphot5a}.

To calculate the EEL and PE probabilities, we expand the incident
electric field due to a moving electron, and also the scattered field
and the fields inside the NP, into vector spherical waves \cite{abajo_prb59,
abajo_rmp82} and apply the boundary conditions of continuity of the
tangential components of the fields at the interfaces between two
different media to obtain the scattering matrix that associates the
amplitude of the scattered field to the incident field. With this
approach the PE and the EEL probabilities can be derived as \cite{abajo_prb59,
matyssek_ultram117}
\begin{equation}\label{Eq:PL}
P_{\mathrm{PE}} = \frac{c^{3}}{4\pi^{2} \omega^{3}}
\sum_{\stackrel{m = -\infty}{l=|m|}}^{\infty}
l \left(l + 1 \right)
\left(\left|F_{lm}\right|^{2}+ \left|G_{lm}\right|^{2} \right),
\end{equation}
\begin{equation}\label{Eq:EEL}
P_{\mathrm{EEL}} = \frac{1}{\pi\omega^{2}}
\sum_{\stackrel{m = -\infty}{l=|m|}}^{\infty}
\left\{ m \nu K_{m}\left(\frac{\omega d}{\nu \gamma}\right) 
\mathrm{Re} \left[(A_{lm}^{+})^{\ast} \mathrm{i} F_{lm}\right] 
+\frac{c}{2\gamma} K_{m}\left(\frac{\omega d}{\nu \gamma}\right) 
\mathrm{Re} \left[(B_{lm}^{+})^{\ast} \mathrm{i} G_{lm}\right] 
\right\}~,
\end{equation}
where ${\mathrm{Re}} [\cdot]$ represents the real part of the function
in square brackets, and the star denotes complex conjugation. In these
expressions $l$ and $m$ are the standard angular momentum indices,  $\gamma=1/\sqrt{1-\nu^2/c^2}$,
$K_{m}$ is the modified Bessel function, and $A_{lm}^{+}$, 
$B_{lm}^{+}$, $F_{lm}$ and $G_{lm}$ are appropriate expansion 
coefficients (see Appendix).

When at least one of the characteristic dimensions of the system
(i.e., $R_{1}$, $R_{2}$, $R_{2}-R_{1}$, or $d$) becomes comparable
to the mean free-electron path, quantum-informed models for the description
of the plasmonic NP response become relevant \cite{zhu_natcom7,
tserkezis_acsphot5b}. Traditional or more advanced hydrodynamic models
are among the most appealing approaches, because they immediately
account for the longer-scale effect of screening, and various implementations
in numerical tools based on boundary elements \cite{trugler_ijmpb31},
finite elements \cite{toscano_oex20}, finite differences in time
domain \cite{mcmahon_prb82}, or discrete sources \cite{eremin_jqsrt217},
have been developed to tackle NPs of arbitrary shapes, while schemes
that can include electron spill-out have also appeared \cite{toscano_natcom6,
ciraci_prb93}. Electron spill-out and tunneling become relevant at
even shorter lengths scales \cite{yan_prl}, and require hybrid models that take as
input fully quantum-mechanical calculations \cite{esteban_natcom3,
hohenester_prb91,yannopapas_ijmpb31,ciraci_nanophot8}. Since, however,
we are interested in exactly solvable analytic solutions, we will resort
here to the standard hydrodynamic Drude model (HDM) \cite{ruppin_prl31}
that accounts for screening, and the generalised nonlocal optical response
(GNOR) theory \cite{mortensen_natcom5} for nonlocal damping. To take
nonlocal effects in the metal into account with these approaches, a
longitudinal term needs to be included in the expansion of the field,
and we employ the standard additional boundary condition of continuity
of the normal component of the displacement field for the no spill-out
case (hard-wall boundary conditions) \cite{tserkezis_nscale8}, which,
despite its simplicity, provides an adequate description of noble metals
like Ag \cite{teperik_oex21}.

\section{Plasmon--exciton coupling in LRA}
\begin{figure}[!t]
\centering
\includegraphics[width=0.70\linewidth]{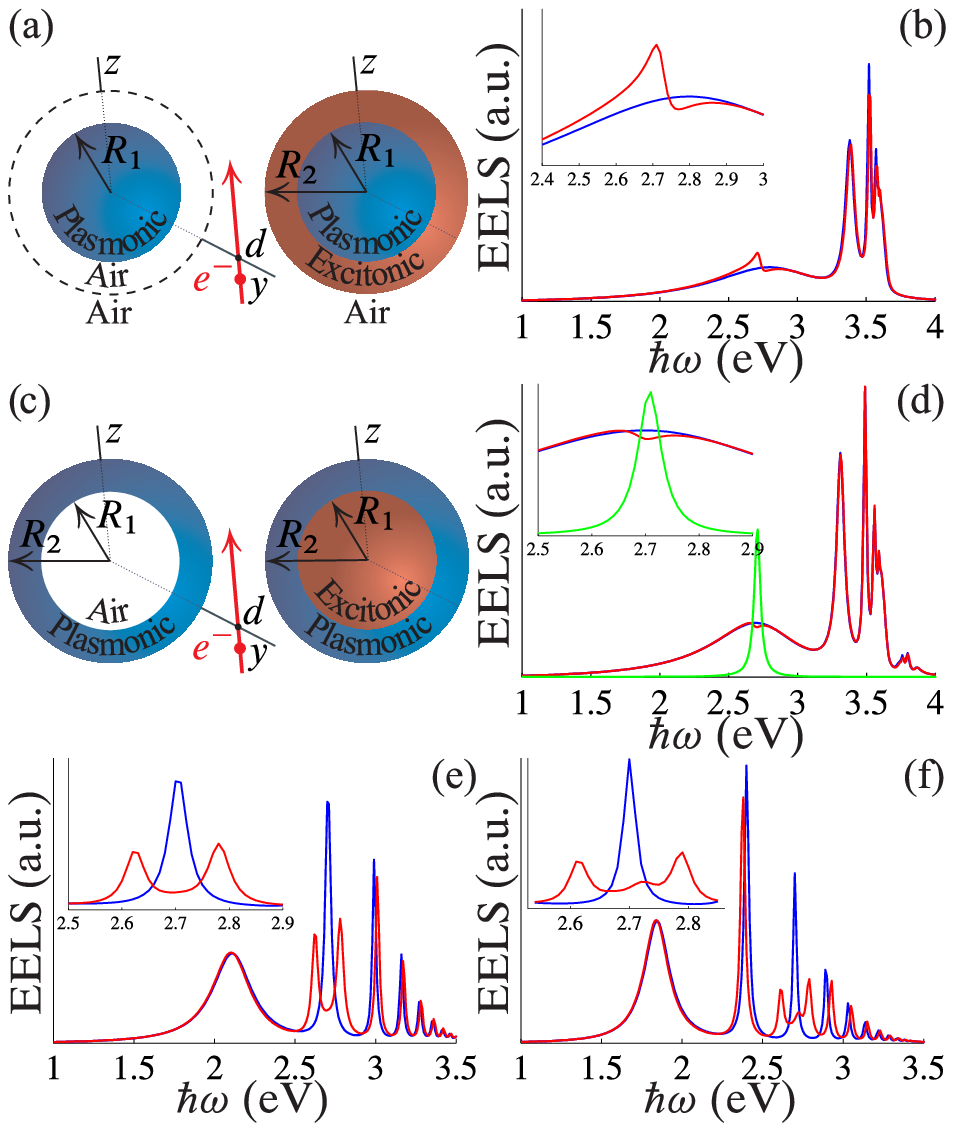}
\caption{
(a) Schematics of the Ag core--exciton shell set-up.
(b) EEL spectra for the set-up of (a), taking $R_{1} = 60$\;nm and
$R_{2} = 65$\;nm. Blue curves correspond to the bare Ag NP spectra
[left-hand schematic in (a)], while red curves correspond to the 
full core--shell NP [right-hand schematic in (a)]. 
(c) Schematics of the exciton core--Ag shell set-up.
(d)--(f) EEL spectra for the set-up of (c), for three different
combinations of radii, so as to align the dipolar, quadrupolar,
and octapolar LSP of the Ag shell to the exciton energy.
In (d) $R_{1} = 33$\;nm and $R_{2} = 60$\;nm, leading to two
hybrid resonances separated by a shallow dip, with separation
of $\hbar \Omega = 0.1$\;eV. In (e) $R_{1} = 51.2$\;nm and
$R_{2} = 60$\;nm. The anticrossing between the quadrupolar Ag
shell LSP and the exciton is $0.16$\;eV. Finally, in (f)
$R_{1} = 54.2$\;nm and $R_{2} = 60$\;nm, and the anticrossing
between the octapolar LSP of the shell and the exciton is
$0.18$\;eV. In all calculations, the impact parameter
is $d = 70$~nm, while the electron speed is $\nu = 0.69 \; c$.
The green curve in (d) corresponds to the bare excitonic NP
spectra [left-hand schematic in (a) where now the core is excitonic
instead of plasmonic]. All insets are zooms in the energy
regions of the relevant resonances.
}\label{Fig1}
\end{figure}

In \fig{Fig1}(b) we show EEL spectra (blue curve) for an Ag sphere
in air (as shown in the left-hand schematics of \fig{Fig1}(a)).
Its radius, $R_{1}$, is set equal to $60$\;nm so that its dipolar LSP
resonance coincides with the $2.7$\;eV resonance energy of the excitonic
material. The dipolar LSP resonance of the Ag sphere manifests in the
EEL spectra as a broad peak at $2.7$\;eV. A smaller NP radius
would blueshift the dipolar LSP (e.g. to $3.14$\;eV for $R_{1} = 40$\;nm),
leading to detuning with $\hbar \omega_{\mathrm{e}}$. Accordingly, larger
radii redshift the LSP modes, thus allowing to fully tune the response.
Adding, next, a $5$\;nm-thick excitonic shell (so that $R_{2} = 65$\;nm),
as shown in the right-hand schematics of \fig{Fig1}(a), has as an
immediate consequence the interaction of the excitonic mode with the
dipolar LSP. Two coupled hybrid modes thus emerge, and their
characteristic anticrossing appears in the spectra, and in standard
resonance energy \textit{vs} detuning diagrams (not shown here).
The spectra of the coupled system are plotted with the red curve
in \fig{Fig1}(b). Clearly, due to the small oscillator strength and
the thinness of the excitonic shell, the two hybrid modes are not
well-discernible; while the first hybrid resonance is well localised at
$2.71$\;eV, the second one, around $2.86$\;nm, is less intense and not
well localised, but almost damped (see inset of \fig{Fig1}(b)).
This is a typical case of weak plasmon--exciton coupling. An attempt
to match the higher-order multipoles of the Ag sphere (appearing in the
EEL spectra at around $3.5$\;eV) to $\hbar \omega_{\mathrm{e}}$ by
further increasing $R_{1}$, results in damped and broadened resonances.
Consequently, the Ag core--exciton shell geometry proves inefficient
in the attempt to achieve a clear Rabi-like splitting for the dipolar
LSP, let alone for higher-order modes, which our intention is to
explore here.

The plasmonic--excitonic configuration of \fig{Fig1}(c) is more
promising for observing the splitting, not only for the dipolar, but
also for the quadrupolar and the octapolar LSP resonances. Such a
nanostructure may be more challenging to fabricate, but on the other
hand protects the excitonic material (especially in case of organic
molecules) from the intense exciting electron beam. Initially,
the thickness $R_{2} - R_{1}$ of the plasmonic shell, in the absence
of the core (left-hand schematic of \fig{Fig1}(c)) is engineered
such that each resonance is tuned to the transition energy of the
excitonic material. Then, by introducing the excitonic core (right-hand
schematic of \fig{Fig1}(c)), plasmon--exciton coupling is allowed
to take place. This is illustrated in \figs{Fig1}(d)--(f) for the
dipolar, quadrupolar and octapolar LSP resonance, respectively. The blue
curves in each figure are the EEL spectra in the absence of the excitonic
core, showing that indeed the mode is tuned to $2.7$\;eV. The red curves
illustrate the splitting in each case, while the insets zoom in the
corresponding spectral window of interest. This set-up reveals
that a clear Rabi-like splitting can be achieved for both higher-order
modes, whilst the dipolar LSP interacts only weakly with the exciton.
This weak interaction, typical of what has been termed induced
transparency region \cite{zengin_scirep3}, could be anticipated by
observing separately the linewidths of the dipolar LSP (Ag shell in 
the absence of the excitonic core) and the excitonic resonance 
(excitonic core in the absence of a shell). The EEL spectrum of this
latter mode is depicted with the green curve in \fig{Fig1}(d).
The linewidth of the excitonic mode is estimated equal to $0.0298$\;eV,
much narrower than the broad linewidth of the LSP. To verify that the
system has indeed entered the strong coupling regime in the case of
the higher-order LSP modes of \figs{Fig1}(e) and (f), we check
whether the observed energy splittings $\hbar \Omega$ satisfy the
strong coupling condition involving the plasmonic ($\gamma_{\mathrm{p}}$) and excitonic
($\gamma_{\mathrm{e}}$)
resonance linewidths \cite{torma_rpp78}
\begin{equation}\label{Eq:condition}
\hbar \Omega >
\left[\frac{(\hbar \gamma_\mathrm{p})^{2}}{2} + 
\frac{(\hbar \gamma_{\mathrm{e}})^{2}}{2} \right]^{1/2}~.     
\end{equation}
In principle, there is also an additional contribution to the broadening from
the interaction with the electron beam, $\gamma_{\mathrm{p-e}}$. Since we are
not strictly interested in displaying a particular strong-coupling architecture,
we will disregard this in what folllows, assuming that $\gamma_\mathrm{p-e} \ll
\gamma_\mathrm{p, rad}$, where $\gamma_\mathrm{p, rad}$ is the radiation
contribution to $\gamma_\mathrm{p}$.
Through Lorentzian fittings for the quadrupolar and octapolar LSPs in
the absence of the excitonic core and in the absence of the Ag shell,
we obtain $\hbar\gamma_{\rm p} = 0.0196$\;eV 
and $\hbar\gamma_{\rm e} = 0.03$\;eV for the case of the quadrupolar mode, and $\hbar\gamma_{\rm p} = 0.0177$\;eV and $\hbar\gamma_{\rm e} = 0.03$\;eV for the case of the octapolar mode. These calculations yield $0.0253$\;eV and $0.0246$\;eV for the quadrupolar and octapolar mode, respectively, as computed by the right-hand side of \eqref{Eq:condition}. On the other hand,
the difference between the two hybrid resonances in \fig{Fig1}(e)
and (f) is $\hbar\Omega = 0.16$\;eV and $\hbar \Omega = 0.18$\;eV, for the
quadrupolar and the octapolar mode respectively, values always greater that the
aforementioned calculated ones from the right-hand side of \eqref{Eq:condition}.
Thus we deduce a distinguishable lower- and higher-order multipolar
strong plasmon--exciton coupling in EELS.

\section{Higher-order plexcitons within quantum-informed models}
Having established that higher-order multipole LSP modes can
efficiently couple with excitons, we go one step further to evaluate
how this coupling is affected by the triggering of quantum effects,
relevant for small NP dimensions (typically below $5$\;nm in radius),
utilising HDM and GNOR as our quantum-informed models. At this scale,
nonlocality plays a decisive role in determining the spectral
features \cite{tserkezis_nscale8}, and it is taken into account
by introducing a compressible electron fluid characterised by a Fermi
velocity $v_{\mathrm{F}} = 1.39 \times 10^{6}$ \; m/s and a diffusion
constant $D = 3.61 \times 10^{-4}$ \; m$^{2}$/s to mimic surface-enhanced Landau damping, values suitable for
Ag \cite{raza_jpcc27} (see also Appendix). Since the
configuration of \fig{Fig1}(a) does not reveal clear anticrossings,
we restrict ourselves to the set-up of \fig{Fig1}(c).

\begin{figure}[!t]
\centering
\includegraphics[width=0.7\linewidth]{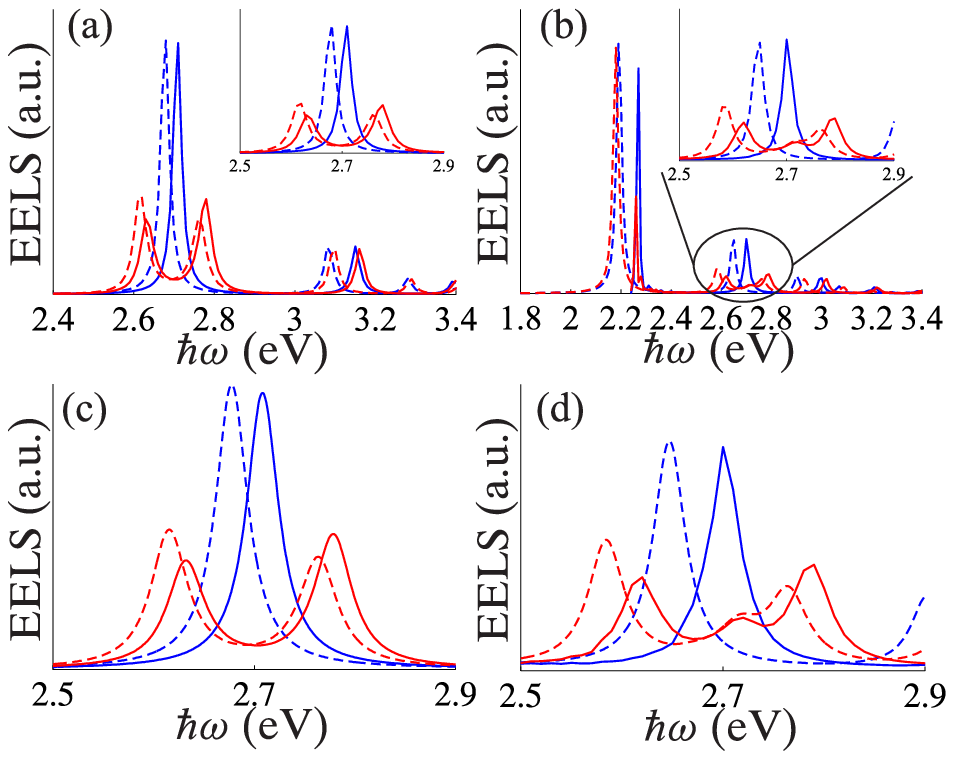}
\caption{Dipolar and quadrupolar higher-order Rabi-like splitting
in plexcitonics as predicted by HDM. In all calculations, the
impact parameter is $d = 5$\;nm and the electron velocity is
$\nu = 0.69 \; c$.
(a) Plasmon--exciton coupling for the dipolar LSP mode, when
$R_{1} = 3.1$\;nm and $R_{2} = 4$\;nm. The blue dashed curve depicts
the EEL spectrum for the set-up of the left-hand schematic of
\fig{Fig1}(c) within LRA, while the blue solid curve shows
the corresponding spectrum within HDM. Red curves correspond
to the presence of the excitonic core, as in the right-hand schematic
of \fig{Fig1}(c). The inset zooms in the energy window of interest.
(b) Quadrupolar LSP--exciton coupling for $R_{1} = 3.51$\;nm and
$R_{2} = 4$\;nm. The blue/red dashed curves depict the EEL spectrum
within LRA without/with the excitonic core and the blue/red solid curves
are the EEL spectrum within HDM without/with the excitonic core present.
(c) EEL spectrum as in the inset of (a). All the values of parameters are the same as in (a) except for $\hbar\gamma_{\rm m}=0.039$~eV in Drude model for Ag.
(d) EEL spectrum as in the inset of (b). All the values of parameters are the same as in (b) except for $\hbar\gamma_{\rm m}=0.039$~eV in Drude model for Ag.
}\label{Fig2}
\end{figure}

We begin our investigation with HDM. The blue dashed and solid curves
in \fig{Fig2}(a) depict EEL spectra for LRA and HDM, respectively,
in the absence of the excitonic core (hollow Ag shell). In this example
the impact parameter is set at $d = 5$\;nm. The electron
velocity is kept at $\nu = 0.69 \; c$, the outer radius is $R_{2} = 4$\;nm,
and a shell thickness of $0.9$\;nm (implying that $R_{1} = 3.1$\;nm) is
such that the LSP dipolar mode of the air--Ag set-up, as calculated within
HDM, matches the transition energy of the excitonic material ($2.7$\;eV).
This is shown by the blue solid curve in \fig{Fig2}(a). With the
same thickness, the LRA dipolar LSP is located at $2.68$\;eV. In these
spectra one can also observe the higher-order (quadrupolar) LRA mode at
$3.08$\;eV, and the corresponding HDM quadrupolar LSP at $3.15$\;eV. Both
dipolar and quadrupolar HDM modes exhibit the anticipated blueshifts as
compared to their LRA counterparts \cite{christensen_nn8}. When the
dye is introduced, plasmon--exciton coupling takes place with the same
$\hbar \Omega = 0.15$\;eV for both LRA and HDM, as illustrated by the
red dashed and solid curves in the inset of \fig{Fig2}(a). Evidently,
nonlocality does not affect the width of the anticrossing, but rather
shifts both hybrid modes by the same amount \cite{tserkezis_acsphot5a},
as one can observe in the inset of \fig{Fig2}(a). Interestingly,
unlike \fig{Fig1}, here strong coupling with the dipolar LSP mode
can be achieved, as the dimensions of the dye layer are such that the
excitonic resonance is comparable in strength. Indeed, the uncoupled
plasmonic and excitonic modes exhibit narrow linewidths
($\hbar \gamma_{\rm p} = 0.0119$\;eV, $\hbar\gamma_{\rm e}=0.0247$\;eV,
as calculated within HDM). These values yield a collective linewidth
(what enters the right-hand side of \eqref{Eq:condition}) of
$0.019$\;eV, much less than the observed $\hbar \Omega = 0.15$\;eV,
thus satisfying the strong coupling condition of \eqref{Eq:condition}.
In \fig{Fig2}(b) we re-engineer the hollow Ag shell so that now
the HDM quadrupolar LSP is tuned to $2.7$\;eV (blue solid curve),
by setting a $0.49$\;nm thickness, while keeping $R_{2} = 4$\;nm. With
this thickness, the LRA quadrupolar LSP appears at $2.65$\;eV (blue
dashed curve in the inset of \fig{Fig2}(b), with $\hbar
\gamma_{\rm p} = 0.0103$\;eV). Introducing the excitonic core
($\hbar \gamma_{\rm e} = 0.0247$\;eV), higher-order Rabi-like
splitting is observed, as shown in \fig{Fig2}(b) by the red
curves, with $\hbar \Omega = 0.17$\;eV. Furthermore, we note that
the slight increase of the radius of the shell, from $3.1$\;nm to
$3.51$\;nm does not affect the exciton linewidth. Ultimately, the
blueshift behaviour due to HDM is inherited by the higher-order
hybrid modes. In \mbox{\figs{Fig2}(c)} and (d) we repeat the study of
\mbox{\figs{Fig2}(a)} and (b), respectively, by increasing the $\hbar\gamma_{\rm m}$
value used in Drude model from $0.025$~eV \mbox{\cite{raza_jpcc27}} to $0.039$~eV
\mbox{\cite{yang_prb91}}, to see how an increased classical damping that might be
experimentally relevant, affects the spectra. As it is evident, the system
still enters the strong coupling regime, with $\hbar\Omega=0.146$~eV,
$\hbar\gamma_{\rm p}=0.0415$~eV, and $\hbar\gamma_{\rm e}=0.0247$~eV
for the dipolar HDM mode [\mbox{\fig{Fig2}(c)}],  and $\hbar\gamma_{\rm p}=0.0379$~eV,
$\hbar\gamma_{\rm e}=0.0247$~eV, $\hbar\Omega=0.169$~eV for the quadrupolar modes of
\mbox{\fig{Fig2}(d)}.

\begin{figure}[!t]
\centering
\includegraphics[width=0.7\linewidth]{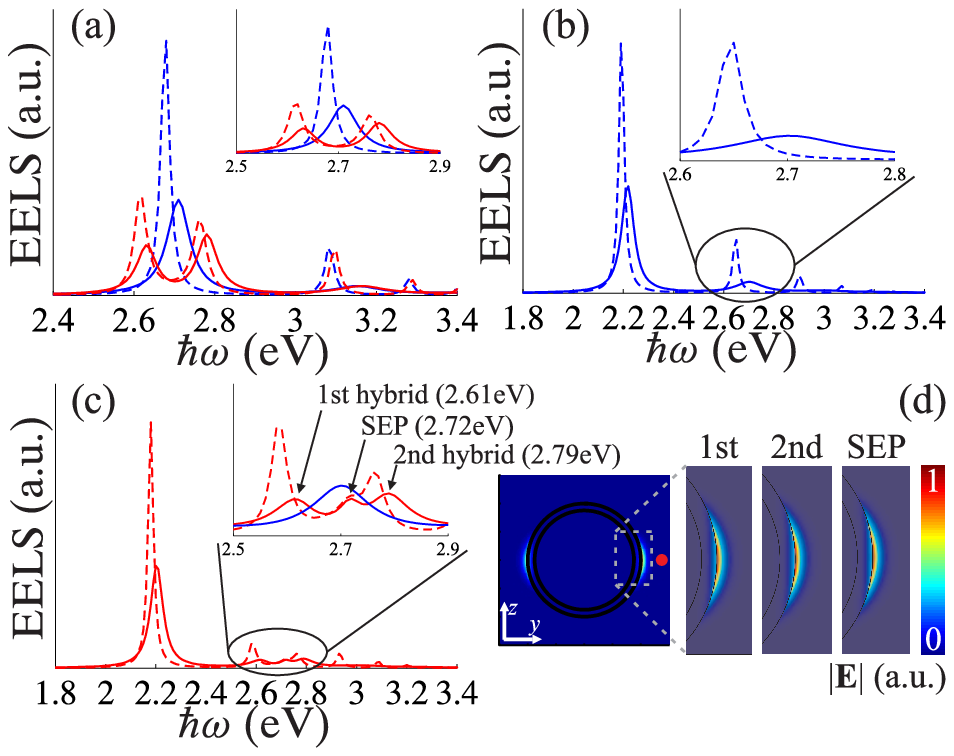}
\caption{Dipolar and higher-order Rabi-like splitting in plexcitonics
as obtained within GNOR. In all calculations, the impact parameter
$d$ and the electron velocity $\nu$ are the same with those in
\fig{Fig2}.
(a) Plasmon--exciton coupling for the dipolar LSP mode, when $R_{1} =
3.1$\;nm and $R_{2} = 4$\;nm. The blue dashed curve depicts the EEL
spectrum for the set-up of the left-hand schematic of \fig{Fig1}(c)
within LRA, while the blue solid curve shows the corresponding spectrum
within GNOR. Red curves correspond to the presence of the excitonic
core. The inset zooms in the energy window of interest.
(b) The quadrupolar LSP resonances (in the absence of the excitonic core)
for $R_{1} = 3.51$\;nm and $R_{2} = 4$\;nm. The blue dashed curve is the
EEL spectrum obtained within LRA, while the solid curve is the
corresponding result within GNOR.
(c) Red curves are the corresponding to (b) EEL spectra when the excitonic
core is present. The blue solid curve in the inset of (c) is as in (b).
(d) Near-field plots for the two coupled hybrid modes and for the SEP
mode shown in the inset of (c). The red bullet depicts the location of the
moving electron.
}\label{Fig3}
\end{figure}

In \fig{Fig3} we take our nonlocal treatment one step further,
to apply the GNOR theory, which is expected to lead to significant mode
broadening due to surface-enhanced Landau damping \cite{tserkezis_acsphot5b}.
To directly compare with \fig{Fig2}, we maintain the same impact
parameter, electron velocity, and outer NP radius, and only change $R_{1}$
to tune the plasmonic response. The blue dashed and solid curves in
\fig{Fig3}(a) depict EEL spectra for LRA and GNOR, respectively, in
the presence of a hollow core. In this case we set $R_{1} = 3.1$\;nm
so that the GNOR LSP dipolar mode of the hollow Ag set-up is tuned to
$2.7$\;eV, while the quadrupolar LSP is located at $3.16$\;eV. Both
GNOR modes exhibit the anticipated blueshifts, and in addition,
nonlocal damping and broadening \cite{christensen_nn8}. Nevertheless,
this additional surface-enhanced Landau damping does not prevent 
the system from entering the strong coupling regime, once the
excitonic core is introduced, with $\hbar \Omega = 0.15$\;eV (for
both LRA and GNOR models) as illustrated by the red dashed and solid
curves in the inset of \fig{Fig3}(a), since the linewidths of the
uncoupled modes are $\hbar \gamma_{\rm p} = 0.0277$\;eV and $\hbar
\gamma_{\rm e} = 0.0247$\;eV.

Since we are more interested in higher-order multipoles, in
\fig{Fig3}(b) the hollow Ag shell is designed to bring the
GNOR quadrupolar LSP at $2.7$\;eV (see inset), by setting a $R_{1} =
3.51$\;nm. Introducing the excitonic core, strong coupling with the
quadrupolar LSP is observed, as shown in \fig{Fig3}(c),
with $\hbar \Omega = 0.18$\;eV. The corresponding uncoupled mode
linewidths are $\hbar \gamma_{\rm p} = 0.0483$\;eV and $\hbar \gamma_{\rm e}
= 0.0247$\;eV, thus fully satisfying \eqref{Eq:condition}. Nevertheless,
in this case, in addition to the two hybrid exciton-polaritons, an 
intermediate resonance is present at $2.72$\;eV (see also inset of
\fig{Fig3}(c)). To conclude about the nature of this mode, we
resort to the LRA spectrum (red dashed curve in this inset). As it is
evident, the two emerged hybrid modes are subject to blueshifts and
broadening due to nonlocality in the metal. Nevertheless, the middle
resonance is unaffected by nonlocality, implying that it corresponds
to a surface exciton polariton (SEP) mode \cite{gentile_jopt19}. This
mode, attributed to the geometrical resonance of a spherical shell
with a negative permeability \cite{antosiewicz_acsphoton1}, was recently
found to be involved in the dipolar Rabi-like splitting of a
dielectric--plasmonic--excitonic set-up \cite{tserkezis_acsphot5a}.
Here it is observed that it can also exist during higher-order
plasmon--exciton coupling in electron probed systems. \fig{Fig3}(d)
depicts the spatial localisation of the two coupled hybrid modes and the
SEP, on the outer surface of the molecule. These maps show that, by setting
the energy of the swift electron to the respective value of the resonance,
the localised electric field pattern is revealed, even at the close
neighborhood of the source. Of course, in modern electron microscopes
this tuneability is not available (they typically operate at a couple
of high voltages), stressing thus the importance of flexibility provided
by the core--shell geometry. Additionally, all three modes have the same
spatial distribution around the surface of the outer sphere, also observed
in electron probed nanorods \cite{konecna_nn12} and
nanopyramids \cite{yankovich_arxiv2019}.

\begin{figure}[!t]
\centering
\includegraphics[width=0.7\linewidth]{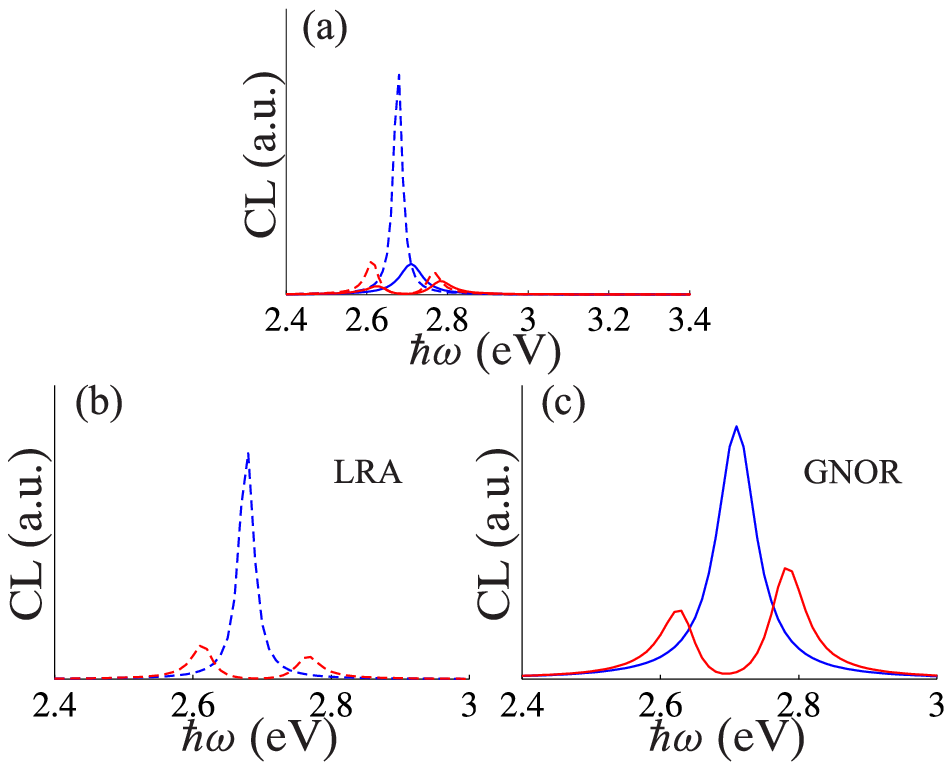}
\caption{CL spectra and dipolar Rabi-like splitting in plexcitonics,
as obtained within LRA and GNOR. All parameters are the same as in
\fig{Fig3}.
(a) Plasmon--exciton coupling for the dipolar LSP mode. Dashed curves
correspond to PE spectra obtained within LRA, while solid curves are
the results of the GNOR model. Additionally, blue curves correspond to
the PE spectra of the hollow Ag shell, and red curves to those of the
excitonic-Ag NP, as in the right-hand schematic of \fig{Fig1}(c).
In (b) and (c) we show an enlarged view of the spectra for the LRA
and GNOR case, respectively.
}\label{Fig4}
\end{figure}

Apart from EELS, CL spectroscopy is also attracting more and more attention
in nanophotonics \cite{vanwijngaarden_apl88,chaturvedi_nn3,gomez_njp10,
jeannin_oex25}. To establish CL as an equally powerful tool for the study
of strong coupling and quantum plasmonics at the same time, at least
when radiative modes are involved, we proceed to calculate PE probabilities
within both classical and hydrodynamic frameworks. In \fig{Fig4}(a) we
plot PE spectra for the same core--shell NPs as in \fig{Fig3}. Comparing
the PE and EEL spectra of the hollow Ag shell (\fig{Fig4}(a) and \fig{Fig3}(a),
respectively) one immediately sees that the dipolar LSP resonance can be clearly
observed in the CL spectra, both for LRA and GNOR, but the quadrupolar LSP
resonance is a predominantly dark mode, as expected. Nevertheless, for the
dipolar LSP both the strongest damping and the blueshift inherent in GNOR
appear in the CL spectra. This is true for both set-ups of the hollow
Ag shell and of the solid excitonic--Ag NP. \figs{Fig4}(b) and (c) depict
separate zoom-ins for LRA and GNOR, once the excitonic core has been introduced.
The anticipated Rabi-like splitting for the dipolar LSP is observable in
CL spectra as well, implying that in the case of radiative modes, CL and
EELS can act complementary to each other, and as efficient substitutes
for optical microscopies. For real experiments, the energy resolution and instrument broadening should naturally be considered, suggesting some advantages of CL over EELS \cite{raza_natcom6}.

\section{Discussion and conclusion}
In summary, we have developed analytic solutions for the EEL and PE
probabilities of core-shell NPs in the presence of nonlocal effects,
taken into account in the general framework of hydrodynamic models.
Applying this formulation to complex core--shell NPs combining a plasmonic
and an excitonic component, we showed that EELS and CL are suitable
complementary techniques to study strong plasmon--exciton coupling.
Focusing in higher-order multipolar LSPs, we showed that it is in
principle feasible to achieve strong coupling, and electron microscopies
offer a more sensitive means to observe this behaviour.

In the realm of strong plasmon-exciton coupling we have discussed how
nonlocal response, as compared to the standard LRA, plays a decisive role in
designing the system and determining the spectral features of the involved
modes. While nonlocality does not affect the width of the anticrossing, but
merely blueshifts both hybrid modes by the same amount, the use of quantum-informed
models is necessary when engineering the plasmonic system and choosing the
excitonic material. The two components need to be accurately tuned to achieve
strong coupling, and the most detailed theoretical predictions of their response
can minimize this effort.

In addition to the resonance positions, the more elaborate GNOR
model contains additional damping mechanisms. Taking these loss channels
into account is fundamental before initiating a quest for strong coupling, 
as there always lurks the risk that coherent energy exchange between the
plasmon and the exciton might prove too slow, and be overcome by absorptive
losses, thus preventing entering the strong coupling regime. Nevertheless,
we have shown that in the examples studied here, nonlocal damping does not
constitute a hindrance, and few-nm NPs could indeed be
considered as candidates for electron microscopy-monitored strong coupling.

Concluding, our analytic work should act as a benchmark for the design
and theoretical study of more elaborate architectures, while our results
should offer further supporting argumentation for turning electron microscopy
into a standard tool in the study of strong coupling.

\appendix*
\section{Derivation of the PE and EEL probabilities in the presence
of nonlocal response for core-shell nanospheres}
\begin{figure}[!t]
\centering
\includegraphics[scale=1.2]{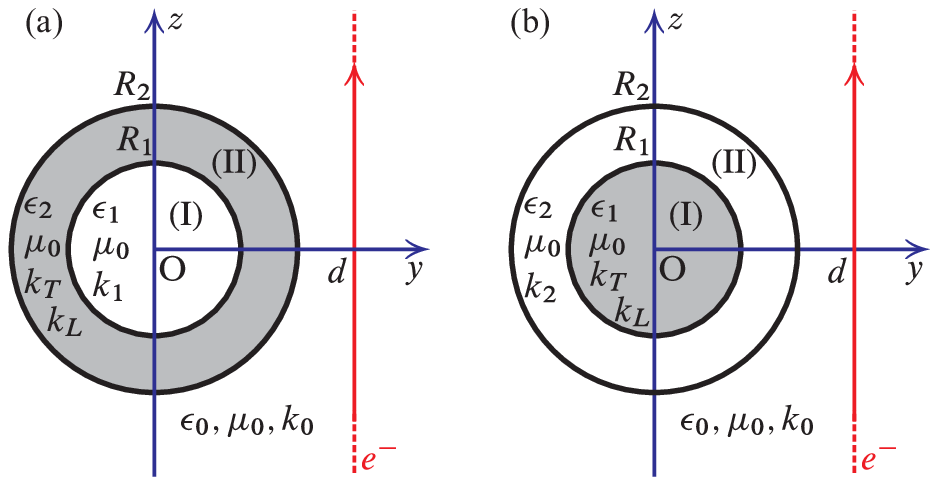}
\caption{Set-up of core-shell nanospheres probed by electrons. (a) dielectric-metallic configuration. (b) metallic-dielectric configuration.}
\label{setup}
\end{figure}

\fig{setup} depicts two set-ups of the core-shell nanoparticle probed by a moving electron. The spherical core has radius $R_1$ and the cladding has outer radius $R_2$. In both cases, the electron has speed $\nu$ and travels at a distance $d$---the impact parameter---from sphere's center. The analysis of CL and EEL response due to the two set-ups, enable us to calculate the PE and EEL probabilities. The two set-ups are examined separately.

\paragraph{Dielectric-metallic nanosphere.}
Referring to \fig{setup}(left), the incident electric field due to a moving electron can be expanded as \cite{abajo_prb59,abajo_rmp82}
\balg{1}
\E^{\rm inc}(\x)=\sum_{\stackrel{\text{\scriptsize$m=-\infty$}}{l=|m|}}^\infty\Big[a_{lm}\mathbf{M}_{lm}^{(1)}(k_0,\x)+b_{lm}\mathbf{N}_{lm}^{(1)}(k_0,\x)\Big],
\ealg
where $\mathbf{M}_{lm}^{(1)}$, $\mathbf{N}_{lm}^{(1)}$ are the spherical vector wave functions (SVWFs) of the first kind \cite{chew}, $k_0=\omega\sqrt{\epsilon_0\mu_0}$ the free space wavenumber, with $\epsilon_0$ and $\mu_0$ the free space permittivity and permeability, respectively. Using the $\exp({\rm i}\omega t)$ time dependence, the expansion coefficients in \eqref{1} are given by
\balg{2}
a_{lm}&=\frac{4\pi {\rm i}\omega\nu}{c^3}\frac{mA_{lm}^+}{l(l+1)}K_m\Big(\frac{\omega d}{\nu\gamma}\Big),\notag\\
b_{lm}&=\frac{2\pi {\rm i}\omega}{c^2\gamma}\frac{B_{lm}^+}{l(l+1)}K_m\Big(\frac{\omega d}{\nu\gamma}\Big),\notag\\
A_{lm}^+&=\sqrt{\frac{2l+1}{\pi}\frac{(l-|m|)!}{(l+|m|)!}}(2|m|-1)!!\Big(\frac{c}{\nu\gamma}\Big)^{|m|} C_{l-|m|}^{|m|+1/2}\Big(\frac{c}{\nu}\Big),\notag\\
B_{lm}^+&=A_{l,m+1}^+\sqrt{(l+m+1)(l-m)}-A_{l,m-1}^+\sqrt{(l-m+1)(l+m)}.
\ealg
In \eqref{2}, $\gamma=1/\sqrt{1-\nu^2/c^2}$, $c$ is the speed of light in vacuum, $K_m$ the modified Bessel function, and $C_{n-|m|}^{|m|+1/2}$ the Gegenbauer polynomial. The expansions of the scattered field $\E^{\rm s}$ and the field $\E^{\rm I}$ inside the dielectric core (region I), are given by 
\balg{3}
\E^{\rm s}(\x)=\sum_{\stackrel{\text{\scriptsize$m=-\infty$}}{l=|m|}}^\infty\Big[F_{lm}\mathbf{M}_{lm}^{(4)}(k_0,\x)+G_{lm}\mathbf{N}_{lm}^{(4)}(k_0,\x)\Big],\notag\\
\E^{\rm I}(\x)=\sum_{\stackrel{\text{\scriptsize$m=-\infty$}}{l=|m|}}^\infty\Big[A_{lm}\mathbf{M}_{lm}^{(1)}(k_1,\x)+B_{lm}\mathbf{N}_{lm}^{(1)}(k_1,\x)\Big],
\ealg
with $\mathbf{M}_{lm}^{(4)}$, $\mathbf{N}_{lm}^{(4)}$ the SVWFs of the fourth kind, representing outward travelling waves, while $k_1=\omega\sqrt{\epsilon_1\mu_0}$, $\epsilon_1=\epsilon_{1r}\epsilon_0$, with $\epsilon_{1r}$ the relative permittivity of the dielectric core. To account for the nonlocal response due to the metallic shell (region II), the field $\E^{\rm II}$ must be expanded taking into account longitudinal waves via the $\mathbf{L}_{lm}$ SVWF, i.e.,
\balg{4}
\!\!\!\E^{\rm II}(\x)=\!\!\!\!\!\sum_{\stackrel{\text{\scriptsize$m=-\infty$}}{l=|m|}}^\infty\Big[&C_{lm}^{(1)}\mathbf{M}_{lm}^{(1)}(k_T,\x)+C_{lm}^{(2)}\mathbf{M}_{lm}^{(2)}(k_T,\x)\notag\\
&+D_{lm}^{(1)}\mathbf{N}_{lm}^{(1)}(k_T,\x)+D_{lm}^{(2)}\mathbf{N}_{lm}^{(2)}(k_T,\x)\notag\\
&+E_{lm}^{(1)}\mathbf{L}_{lm}^{(1)}(k_L,\x)+E_{lm}^{(2)}\mathbf{L}_{lm}^{(2)}(k_L,\x)\Big].
\ealg
In \eqref{4}, $\mathbf{M}_{lm}^{(2)}$, $\mathbf{N}_{lm}^{(2)}$, $\mathbf{L}_{lm}^{(2)}$ are the SVWFs of the second kind. $k_T=\omega\sqrt{\epsilon_2\mu_0}$, with $\epsilon_2=\epsilon_{2r}\epsilon_0$, is the transverse wavenumber of the LRA of the metallic shell, with the relative dielectric function $\epsilon_{2r}$ following Drude model
\balg{5}
\epsilon_{2r}(\omega)=\epsilon_\infty(\omega)-\frac{\omega_{\rm p}^2}{\omega(\omega-{\rm i}\gamma_{\rm m})}.
\ealg
Here, $\epsilon_\infty$ accounts for interband effects, $\omega_{\rm p}$ is the plasma frequency, and $\gamma_{\rm m}$ is the damping rate. The longitudinal wavenumber $k_L=\sqrt{\epsilon_{2r}(\omega)}/\xi(\omega)$ depends on the model of nonlocality, either the hydrodynamic or the GNOR model \cite{raza_jpcc27}. For the HDM, $\xi(\omega)\equiv\xi_{\rm H}(\omega)=\beta\sqrt{\epsilon_\infty(\omega)}/\sqrt{\omega(\omega-{\rm i}\gamma_{\rm m})}$, with the hydrodynamic parameter $\beta=\sqrt{1/3}v_{\rm F}$ when $\omega\ll\gamma_{\rm m}$, and $\beta=\sqrt{3/5}v_{\rm F}$ when $\omega\gg\gamma_{\rm m}$. In the two latter expressions, $v_{\rm F}$ is Fermi velocity. For the GNOR model, $\xi(\omega)\equiv\xi_{\rm GNOR}(\omega)=\sqrt{\epsilon_\infty(\omega)}\sqrt{\beta^2+D(\gamma_{\rm m}+{\rm i}\omega)}/\sqrt{\omega(\omega-{\rm i}\gamma_{\rm m})}$, where $D$ is the diffusion constant. Values of $v_{\rm F}$ and $D$ are tabulated for various plasmonic metals \cite{bla_arn_for_09,raza_jpcc27,yang_prb91}. Corresponding expansions for the magnetic fields $\H^{\rm inc}$, $\H^{\rm s}$, $\H^{\rm I}$, $\H^{\rm II}$, are obtained by $\H=-\nabla\times\E/({\rm i}\omega\mu)$. It is important to note that the magnetic fields do not feature the $\mathbf{L}_{lm}$ SVWF, since $\nabla\times\mathbf{L}_{lm}=0$.

Matching the boundary conditions $\ur\times(\E^{\rm II}-\E^{\rm I})=0$, $\ur\times(\H^{\rm II}-\H^{\rm I})=0$ on inner surface $r=R_1$, $\ur\times(\E^{\rm inc}+\E^{\rm s}-\E^{\rm II})=0$, $\ur\times(\H^{\rm inc}+\H^{\rm s}-\H^{\rm II})=0$ on outer surface $r=R_2$, as well as the additional boundary conditions $\ur\cdot(\epsilon_\infty\epsilon_0\E^{\rm II}-\epsilon_{1r}\epsilon_0\E^{\rm I})=0$ on $r=R_1$ and $\ur\cdot(\epsilon_0\E^{\rm inc}+\epsilon_0\E^{\rm s}-\epsilon_\infty\epsilon_0\E^{\rm II})=0$ on $r=R_2$, to account for the nonlocal effects, we get two separate linear systems for the calculation of the unknown expansion coefficients appearing in \eqrefs{3} and \eqrefnt{4}. The first system reads
\balg{6}
\mathbb{A}
\begin{bmatrix}
C_{lm}^{(1)}\\
C_{lm}^{(2)}\\
A_{lm}\\
F_{lm}
\end{bmatrix}
=
\begin{bmatrix}
0\\
j_l(k_0R_2)a_{lm}\\
0\\
j_l^{\rm d}(k_0R_2)/(Z_0k_0R_2)a_{lm}
\end{bmatrix},
\ealg
with $\mathbb{A}=(a_{ij})$, $i,j=1,\ldots,4$, and
\balg{Ael}
a_{11}&=j_l(k_TR_1),a_{12}=y_l(k_TR_1),a_{13}=-j_l(k_1R_1),a_{14}=0,\notag\\
a_{21}&=j_l(k_TR_2),a_{22}=y_l(k_TR_2),a_{23}=0,a_{24}=-h_l^{(2)}(k_0R_2),\notag\\
a_{31}&=\frac{j_l^{\rm d}(k_TR_1)}{Z_2k_TR_1},a_{32}=\frac{y_l^{\rm d}(k_TR_1)}{Z_2k_TR_1},a_{33}=-\frac{j_l^{\rm d}(k_1R_1)}{Z_1k_1R_1},a_{34}=0,\notag\\
a_{41}&=\frac{j_l^{\rm d}(k_TR_2)}{Z_2k_TR_2},a_{42}=\frac{y_l^{\rm d}(k_TR_2)}{Z_2k_TR_2},a_{43}=0, a_{44}=-\frac{h_l^{(2){\rm d}}(k_0R_2)}{Z_0k_0R_2}.
\ealg
In above relations, $j_l$, $y_l$, $h_l^{(2)}$ is the spherical Bessel, Neumann, Hankel function of the second kind, $z_l^{\rm d}(x)\equiv[xz_l(x)]'_x$, $z_l\equiv j_l,y_l,h_l^{(2)}$, and $Z_{0,1,2}=\sqrt{\mu_{0}/\epsilon_{0,1,2}}$. The second linear system is given by
\balg{7}
\mathbb{B}
\begin{bmatrix}
D_{lm}^{(1)}\\
D_{lm}^{(2)}\\
E_{lm}^{(1)}\\
E_{lm}^{(2)}\\
B_{lm}\\
G_{lm}
\end{bmatrix}
=
\begin{bmatrix}
0\\
j_l^{\rm d}(k_0R_2)/(k_0R_2)b_{lm}\\
0\\
j_l(k_0R_2)/Z_0b_{lm}\\
0\\
l(l+1)j_l(k_0R_2)/(k_0R_2)b_{lm}
\end{bmatrix},
\ealg
with $\mathbb{B}=(b_{ij})$, $i,j=1,\ldots,6$, while
\balg{Bel}
b_{11}&=\frac{j_l^{\rm d}(k_TR_1)}{k_TR_1},b_{12}=\frac{y_l^{\rm d}(k_TR_1)}{k_TR_1}, b_{13}=\frac{j_l(k_LR_1)}{k_LR_1}, b_{14}=\frac{y_l(k_LR_1)}{k_LR_1}, b_{15}=-\frac{j_l^{\rm d}(k_1R_1)}{k_1R_1}, b_{16}=0,\notag\\
b_{21}&=\frac{j_l^{\rm d}(k_TR_2)}{k_TR_2}, b_{22}=\frac{y_l^{\rm d}(k_TR_2)}{k_TR_2}, b_{23}=\frac{j_l(k_LR_2)}{k_LR_2}, b_{24}=\frac{y_l(k_LR_2)}{k_LR_2}, b_{25}=0,\notag\\ b_{26}&=-\frac{h_l^{(2){\rm d}}(k_0R_2)}{k_0R_2},\notag\\
b_{31}&=\frac{j_l(k_TR_1)}{Z_2}, b_{32}=\frac{y_l(k_TR_1)}{Z_2}, b_{33}=0, b_{34}=0, b_{35}=-\frac{j_l(k_1R_1)}{Z_1}, b_{36}=0,\notag\\
b_{41}&=\frac{j_l(k_TR_2)}{Z_2}, b_{42}=\frac{y_l(k_TR_2)}{Z_2}, b_{43}=0, b_{44}=0, b_{45}=0, b_{46}=-\frac{h_l^{(2)}(k_0R_2)}{Z_0},\notag\\
b_{51}&=\epsilon_\infty l(l+1)\frac{j_l(k_TR_1)}{k_TR_1}, b_{52}=\epsilon_\infty l(l+1)\frac{y_l(k_TR_1)}{k_TR_1}, b_{53}=\epsilon_\infty j'_l(k_LR_1), b_{54}=\epsilon_\infty y'_l(k_LR_1),\notag\\
b_{55}&=-\epsilon_{1r} l(l+1)\frac{j_l(k_1R_1)}{k_1R_1}, b_{56}=0,\notag\\
b_{61}&=\epsilon_\infty l(l+1)\frac{j_l(k_TR_2)}{k_TR_2}, b_{62}=\epsilon_\infty l(l+1)\frac{y_l(k_TR_2)}{k_TR_2}, b_{63}=\epsilon_\infty j'_l(k_LR_2), b_{64}=\epsilon_\infty y'_l(k_LR_2),\notag\\
b_{65}&=0, b_{66}=-l(l+1)\frac{h_l^{(2)}(k_0R_2)}{k_0R_2}.
\ealg
The prime appearing in $j'_l$, $y'_l$ denotes differentiation with respect to the argument.

Once the expansion coefficients of the scattered field are known from the solution of \eqrefs{6} and \eqrefnt{7}, the PE probability can be evaluated by \cite{abajo_prb59,mat_sch_her_wri_12}
\balg{8}
P_{\rm PE}=\frac{c^3}{4\pi^2\omega^3}\sum_{\stackrel{\text{\scriptsize$m=-\infty$}}{l=|m|}}^\infty l(l+1)(|F_{lm}|^2+|G_{lm}|^2),
\ealg
and the EEL probability by \cite{abajo_prb59}
\balg{9}
P_{\rm EEL}&=\frac{1}{\pi\omega^2}\sum_{\stackrel{\text{\scriptsize$m=-\infty$}}{l=|m|}}^\infty \Big\{m\nu K_m\Big(\frac{\omega d}{\nu\gamma}\Big){\rm Re}[(A_{lm}^+)^\ast {\rm i}F_{lm}]+\frac{c}{2\gamma}K_m\Big(\frac{\omega d}{\nu\gamma}\Big){\rm Re}[(B_{lm}^+)^\ast {\rm i}G_{lm}]\Big\},
\ealg
where ${\rm Re}[\cdot]$ represents the real part, and the star denotes complex conjugation.

\paragraph{Metallic-dielectric nanosphere.}
In this case, \eqref{1} and $\E^{\rm sc}$ in \eqref{3} remain the same, though $\E^{\rm I}$ and $\E^{\rm II}$ must be expanded as
\balg{10}
\!\!\!\!\!\E^{\rm I}(\x)=\!\!\!\!\!\sum_{\stackrel{\text{\scriptsize$m=-\infty$}}{l=|m|}}^\infty\Big[&A_{lm}\mathbf{M}_{lm}^{(1)}(k_T,\x)+B_{lm}\mathbf{N}_{lm}^{(1)}(k_T,\x)+C_{lm}\mathbf{L}_{lm}^{(1)}(k_L,\x)\Big],\notag\\
\!\!\!\!\!\E^{\rm II}(\x)=\!\!\!\!\!\sum_{\stackrel{\text{\scriptsize$m=-\infty$}}{l=|m|}}^\infty\Big[&D_{lm}^{(1)}\mathbf{M}_{lm}^{(1)}(k_2,\x)+D_{lm}^{(2)}\mathbf{M}_{lm}^{(2)}(k_2,\x)+E_{lm}^{(1)}\mathbf{N}_{lm}^{(1)}(k_2,\x)+E_{lm}^{(2)}\mathbf{N}_{lm}^{(2)}(k_2,\x)\Big].
\ealg
Now $k_T=\omega\sqrt{\epsilon_1\mu_0}$, $\epsilon_1=\epsilon_{1r}\epsilon_0$, where $\epsilon_{1r}$ represents the relative permittivity of the plasmonic core and is again given by Drude model of \eqref{5}. Furthermore, $k_L=\sqrt{\epsilon_{1r}(\omega)}/\xi(\omega)$, and $k_2=\omega\sqrt{\epsilon_2\mu_0}$, $\epsilon_2=\epsilon_{2r}\epsilon_0$, with $\epsilon_{2r}$ the relative permittivity of the dielectric coating.

Satisfying the additional boundary condition $\ur\cdot(\epsilon_{2r}\epsilon_0\E^{\rm II}-\epsilon_\infty\epsilon_0\E^{\rm I})=0$ on $r=R_1$, as well as the remaining boundary conditions for the continuity of the transversal field components, we again get two separate linear systems for the determination of the unknown coefficients. The first system is the same with \eqref{6}, but the unknown vector is now $[D_{lm}^{(1)},\;D_{lm}^{(2)},\;A_{lm},\;F_{lm}]^T$, whilst $k_T$ and $k_1$ appearing in $a_{ij}$ of \eqref{6}, must be substituted by $k_2$ and $k_T$, respectively. The second system is now given by
\balg{11}
\mathbb{B}
\begin{bmatrix}
E_{lm}^{(1)}\\
E_{lm}^{(2)}\\
B_{lm}\\
C_{lm}\\
G_{lm}
\end{bmatrix}
=
\begin{bmatrix}
0\\
j_l^{\rm d}(k_0R_2)/(k_0R_2)b_{lm}\\
0\\
j_l(k_0R_2)/Z_0b_{lm}\\
0
\end{bmatrix},
\ealg
with $\mathbb{B}=(b_{ij})$, $i,j=1,\ldots,5$, and
\balg{BBel}
b_{11}&=\frac{j_l^{\rm d}(k_2R_1)}{k_2R_1}, b_{12}=\frac{y_l^{\rm d}(k_2R_1)}{k_2R_1}, b_{13}=-\frac{j_l^{\rm d}(k_TR_1)}{k_TR_1}, b_{14}=-\frac{j_l(k_LR_1)}{k_LR_1}, b_{15}=0,\notag\\
b_{21}&=\frac{j_l^{\rm d}(k_2R_2)}{k_2R_2}, b_{22}=\frac{y_l^{\rm d}(k_2R_2)}{k_2R_2}, b_{23}=0, b_{24}=0, b_{25}=-\frac{h_l^{(2){\rm d}}(k_0R_2)}{k_0R_2},\notag\\
b_{31}&=\frac{j_l(k_2R_1)}{Z_2}, b_{32}=\frac{y_l(k_2R_1)}{Z_2}, b_{33}=-\frac{j_l(k_TR_1)}{Z_1}, b_{34}=0, b_{35}=0,\notag\\
b_{41}&=\frac{j_l(k_2R_2)}{Z_2}, b_{42}=\frac{y_l(k_2R_2)}{Z_2}, b_{43}=0, b_{44}=0, b_{45}=-\frac{h_l^{(2)}(k_0R_2)}{Z_0},\notag\\
b_{51}&=\epsilon_{2r}l(l+1)\frac{j_l(k_2R_1)}{k_2R_1}, b_{52}=\epsilon_{2r}l(l+1)\frac{y_l(k_2R_1)}{k_2R_1}, b_{53}=-\epsilon_\infty l(l+1) \frac{j_l(k_TR_1)}{k_TR_1},\notag\\
b_{54}&=-\epsilon_\infty j'_l(k_LR_1), b_{55}=0.
\ealg
Then, CL spectra and EEL spectra are obtained via \eqrefs{8} and \eqrefnt{9}, respectively.

\begin{acknowledgments}
We thank Saskia Fiedler for carefully reading and commenting
on the manuscript.
G.~P.~Z. and G.~D.~K. were supported by DAAD program ``Studies on generalized multipole techniques and the method of auxiliary sources, with applications to electron energy loss spectroscopy''.
N.~A.~M. is a VILLUM Investigator supported by VILLUM FONDEN
(grant No. 16498). The Center for Nano Optics is financially supported
by the University of Southern Denmark (SDU 2020 funding).
\end{acknowledgments}


%

\end{document}